**Engineering Sustainability**

# The economic value of transport infrastructure in the UK: an input–output analysis

**Nikolaos Kalyviotis** Dipl-Ing, MSc, MSPM, MBA, PhD
Civil Engineer, Technical Studies Department, Directorate of Technical Works, University of Crete, Rethymno, Greece
(corresponding author: nkalyviotis@uoc.gr)

**Christopher D. F. Rogers** Eur Ing, BSc, PhD, CEng, MICE, FCIHT, FISTT, SFHEA
Professor of Geotechnical Engineering, Department of Civil Engineering, University of Birmingham, Birmingham, UK (Orcid:0000-0002-1693-1999)

**Geoffrey J. D. Hewings** BA, MA, PhD
Emeritus Professor of Urban & Regional Planning, Economics and Geography & Regional Science, Regional Economics Applications Laboratory, University of Illinois at Urbana–Champaign, Urbana, IL, USA (Orcid:0000-0003-2560-3273)

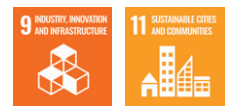

Transport infrastructure systems operate in, and are shaped by, the specific context in which they are expected to perform and contribute to the system-of-systems that support civilised life; they must strive to be sustainable and resilient. However, transport, one of the 'economic infrastructures', is viewed narrowly in political circles as a vehicle for economic prosperity: political focus falls on the economic pillar of sustainability. One perennial challenge concerns the inclusion of social and environmental value into such performance judgements. A deep understanding of system interdependencies is essential when comparing input–output tables (a top-down approach) with cost–benefit analysis (CBA, a bottom-up approach). CBA works best when there is adequate information to calculate a monetary value for all economic, social, and environmental outcomes, whereas input–output tables are most effective when environmental and social value are kept distinct from economic value and where the beneficial economic impacts extend across multiple sectors (CBA is inadequate in this regard). This research uses the World Input-Output Database and principal component analysis to develop a model to capture this complex accounting process. In recognising how past interdependencies inform future development, this study's model provides new insights into indirect economic value creation within infrastructure systems.

**Keywords:** economics & finance/infrastructure planning/transport planning/UN SDG 9: Industry, innovation and infrastructure/UN SDG 11: Sustainable cities and communities

## Notation

| | |
|---|---|
| $X_{cr1}$ | value added from energy sector to transport sector |
| $X_{cr2}$ | value added from waste sector to transport sector |
| $X_{cr3}$ | value added from communication sector to transport sector |
| $X_{cr4}$ | value added from water sector to transport sector |
| $X_{cr5}$ | value added from printing and reproduction of recorded media sector to transport sector |
| $X_{cr6}$ | value added from manufacture of textiles wearing apparel, and leather products sector to transport sector |
| $X_{cr7}$ | value added from mining and quarrying sector to transport sector |
| $Y_{cr}$ | value of transport sector |

## 1. Introduction

There is an ongoing debate about the value of the economic benefits of infrastructure and how to prioritise infrastructure investments in the UK while considering also environmental and social values (iBUILD, 2017). Infrastructure is generally defined as 'a large-scale physical resource made by humans for public consumption' (Frischmann, 2012: p. 3) and is commonly conceptualised in relation to economic interdependencies. Starting from the general interdependency theory, economists call interdependencies 'synergies' (Steinmueller, 1996) and engineers call them 'interconnections' (Hall *et al*., 2016). Traditional management protocols are insufficient for addressing the interconnected nature of modern infrastructure (O'Brien and MacAskill, 2023). The different terms used for dependencies and interdependencies by the social sciences and engineering highlight the need for interdisciplinary research embracing these two, and indeed all other relevant, domains (Hargreaves *et al*., 2020; Leach and Rogers, 2020). A good example concerns the health and well-being benefits of active travel: the former might be approximately monetised in a general and very approximate sense, yet the latter are problematic and, from an engineering provision perspective 'assessing the quality of pedestrian networks and facilities [is a challenge because of] the diverse nature of pedestrians and their sensitivity to subjective influences' (Reid, 2008). The search for a better understanding of the impact of infrastructure interdependencies on transport infrastructure system performance has led to the consideration of new business models, which are defined herein as all positive outcomes balanced against all negative outcomes (including the need for investment).

More generally, there has been a profound shift over recent years towards recognising the need for taking a systems approach to the design, delivery and operation of infrastructure (Rogers *et al*., 2023), of focusing on value rather than cost, and the need for novel business models that account for the full extent and magnitude of value created (World Economic Forum, 2019). This, in turn, requires methods of identifying all the dependencies and interdependencies between a system of interest (here, the transport system) and all other societal and environmental systems (see e.g. Cavada *et al*., 2021). Critically, it is these new business models that enable transformational transport schemes to be implemented, and equally these business models must be themselves resilient to contextual change. The very concept of resilience has been widely explored (e.g. Rogers, 2018; Rogers *et al*., 2012a) and methods of assessing, and hence designing strategies to deliver, resilience have been developed (Hunt *et al*., 2013; Lombardi *et al*., 2012; Rogers, 2010; Rogers *et al*., 2012b). Incorporating resilience involves assessing the ability of infrastructure to withstand and recover from adverse contextual changes, whether rapid (such as natural disasters and economic shocks) or slowly progressing (such as demographic change and increasing demands on the system).

For all of the above reasons, resilience is increasingly recognised as a critical factor in the economic evaluation of transport infrastructure (Marsland *et al*., 2022). This approach ensures that investments are not only economically viable but also sustainable and robust in the face of uncertainties. For instance, the World Bank emphasises the importance of resilience in transport infrastructure projects, highlighting that resilient infrastructure can lead to significant economic savings and enhanced social welfare (World Bank, 2019), noting that the latter has an evident indirect economic value.

Moreover, the value of nature should be rethought, considering not just financial but also economic, environmental, and social values (Ashley *et al*., 2023). This paper therefore focuses on addressing the economic, social, and environmental values of engineering with sustainability in mind and the benefit or otherwise of monetisation to support system changes. It acknowledges that engineering solutions need to be both sustainable and resilient to tackle the challenges presented by increasingly complex and interdependent infrastructures (Hojjati *et al*., 2018) and that building in both of these principles manifestly adds value and extends the scope of investment decisions. A further influence concerns political decision making, for example by influencing climate change responses, which in turn affect infrastructure investments (Mell *et al*., 2022). This multidimensional conceptualisation of business models underscores the necessity for interdisciplinary research to identify and fully characterise infrastructure interdependencies for what is ultimately a holistic approach to sustainability. While identification and formulation of narratives to describe interdependencies is possible using the methodology outlined by Cavada *et al*. (2021) and Rogers *et al*. (2023), the quest for their adequate quantification remains a challenge; an economic equivalency provides one solution for which input–output analysis (a top-down approach) and cost–benefit analysis (CBA, a bottom-up approach) can be applied. The criticality of comprehensively accounting for all value gained or compromised (wherever this value arises) for infrastructure investment prioritisation has significant implications for policy-making and corporate responsibility, and is matched by the allied benefit of de-risking the investment by revealing all the likely outcomes of the system change (Rogers *et al*., 2023). This paper explores the best way of achieving this goal.

## 2. Literature review

### 2.1 Defining infrastructure and the existing economic environment

The notion of infrastructure as 'a large-scale physical resource made by humans for public consumption' (Frischmann, 2012) and 'capital goods that are not directly consumed and serve as support to the functions of a society (individuals, institutions, and corporations)' (Rodrigue, 2017) positions the general public as the users and, indirectly, advisors to the client on the need and form required. This conjures up the idea of a straightforward transaction: paying for economic, societal, and environmental outcomes (Leach *et al*., 2020), accounted for by a return on investment. Monetisation of the beneficial outcomes is a logical action. Transport, communication, water, waste, and energy systems are examples of the large-scale physical resources essential for the functioning of any society. Collectively known as 'economic infrastructures', their definition has evolved over time (see Table 1), reflecting changes in society.

The different definitions emphasise that infrastructures provide network services and public goods (Grimsey and Lewis, 2002; Jessen, 1984; Martini and Lee, 1996; National Science Foundation, 2017; Pearlstein, 2014), interact and interrelate with the socioeconomic system (Allenby and Chester, 2018; Oughton *et al*., 2018; Sussman *et al*., 2009), and influence the social value in terms of how individuals perceive and comprehend the value of each infrastructure (Oughton *et al*., 2018). Allenby and Chester's definition differs from the others and is less clear in its meaning, as it implies that infrastructure is not only a human construct. Allenby and Chester's definition stems from an academic discussion from an environmental perspective, as this definition attempts to incorporate both human-built infrastructure and the natural environment and the resources it provides in a single (planetary) system – an important reminder to engineers that what we create should augment the ecosystem services that the planet provides rather than seeking to replace them. Thus, infrastructure provision was defined based on its physical components initially and then the socioeconomic value was incorporated into infrastructure's definition, while current infrastructure definitions embrace also environmental value.

This research adopts the definition by Frischmann (2012) by examining economic, social, and environmental value dependencies of infrastructure systems from both engineering and economic perspectives. This is distinct from definitions that either concentrate on infrastructure projects (Jessen, 1984; Martini and Lee, 1996) or on economic and policy aspects (National Science Foundation, 2017; Pearlstein, 2014; Sussman *et al.*, 2009) or depict new economic environments and approaches (Allenby and Chester, 2018; Oughton *et al.*, 2018). The services/sectors of interest to this study are transport, waste, water, energy, and communication (Hall *et al.*, 2016:

p. 10; iBUILD, 2017). Every infrastructure system has a network (lines, roads, canals, and so on), some terminals (web servers, airports, treatment plants, and so on), and modes of transfer (cables, pipelines, vehicles, and so on).

Table 1. Civil infrastructure definitions (Kalyviotis, 2022)

| Author | Definition |
| --- | --- |
| Jessen (1984) | Infrastructure as 'public works' including 'roads, bridges, dams, mass transit systems, and sewage and water systems' (Jessen, 1984) |
| Martini and Lee (1996) | Infrastructure provides 'basic services to industry and household' (Martini and Lee, 1996) |
| Sussman *et al.* (2009) | 'Complex, large-scale, interconnected, open, sociotechnical systems' (Sussman *et al.*, 2009: p. 4) |
| Frischmann (2012) | Infrastructure is a 'large-scale physical resource made by humans for public consumption' (Frischmann, 2012: p. 3) |
| Pearlstein (2014) | Infrastructure is a 'large capital intensive natural monopolies such as highways, other transportation facilities, water and sewer lines, communication systems often publicly owned' or 'the tangible capital stock owned by the public sector' |
| National Science Foundation (2017) | 'Infrastructures are defined as networks of systems and processes that function cooperatively and synergistically to produce and distribute a continuous flow of essential goods and services' (National Science Foundation, 2017: p. 14) |
| Oughton *et al.* (2018) | 'Infrastructure is an enabling system that provides a range of different services to intermediate and end users' (Oughton *et al.*, 2018: p. 2) |
| Allenby and Chester (2018) | 'The planet as infrastructure' (Allenby and Chester, 2018) |

Transport, energy, water, waste, communications, and housing provide 'basic infrastructure services' (Menéndez, 1991). The authors are aware that there are other critical types of infrastructure systems, such as human health, education, common defence, and parks, which have significant economic implications and intersect with transport systems. This research focuses on the 'basic infrastructure services' since they have far-reaching impacts on the other types of infrastructure sectors. 'Basic infrastructure services are services that allow the urban poor to live under conditions that facilitate their income-generating activities so they can maintain a good nutritional level and participate in the normal activities of society' (Menéndez, 1991).

The economic environment affects infrastructure and society. The mainstream economic environments of the twentieth century are shareholder and stakeholder capitalism. Shareholder capitalism assumes that corporations are owned by their shareholders and that their goal is to maximise shareholder value (Gompers *et al.*, 2003). This approach has led to high economic growth but also to negative externalities such as social and environmental costs, inequality, and short-termism (Pearlstein, 2014). Stakeholder capitalism considers the well-being of all stakeholders affected by an action or decision, including the shareholders, employees, final users, society, and environment (Freeman *et al.*, 2007). This approach relies on social and environmental values, rather than simply economic value, and is more suitable for civil infrastructure, which serves the long-term needs of society.

While water and waste infrastructures cover basic needs (notably public health) and are essential to support human life, all of the economic infrastructures (transport, water, waste, energy, and communication) serve the long-term needs of society and are essential in providing economic, social, and environmental value to society (Hay, 2016). Given the long-term nature of infrastructure investments, long-term operation and maintenance are required and this also needs to be built into business models and investment decisions.

## 3. Methodology

### 3.1 Methods and assumptions

This paper adopts a large-scale (strategic) analysis of the economic value of infrastructure, which is contingent on the scale of the analysis. The more recent definitions of infrastructure (Allenby and Chester, 2018; Frischmann, 2012; National Science Foundation, 2017; Oughton *et al.*, 2018; Pearlstein, 2014; Sussman *et al.*, 2009) inform this analysis, rather than the earlier ones that regard infrastructure primarily as projects (Jessen, 1984; Martini and Lee, 1996), while adopting the common starting point that the economic value of infrastructure investment determines the feasibility of an infrastructure project.

Today's political and economic systems are largely influenced by the mainstream economic approach of the twentieth century, which is dominated by the concept of a linear economy. Economic value is created by humans through economic exchange and through flows of extraction of raw materials and the production and consumption of goods and services (Stahel and MacArthur, 2019). The economic appraisal of an infrastructure project by a private entity (shareholder capitalism) focuses on the shareholder's economic value (McCann and Berry, 2017). The economic, social, and environmental value for society is the focus of the appraisal of an infrastructure project from a stakeholder viewpoint (stakeholder capitalism). CBA and input–output tables are two different methods widely used to assess the economic value of infrastructure projects.

CBA relies on the extensive available data gathered from previous projects of a similar nature (Hoogmartens *et al*., 2014). The information associated with such projects is theoretically readily available (Van Wee, 2007). CBA evaluates the economic performance of infrastructure in monetary units using different monetary criteria and methods (Vickerman, 2007). CBA attempts to capture social and environmental value by estimating positive and negative externalities over the lifetime of the project (Atkins *et al*., 2017). Although there are many CBA techniques that can be used as a tool for social value measurement (Fujiwara *et al*., 2022), Freelove and Gramatki (2022) point out that CBA cannot always measure and monetise all possible outcomes of large projects and emphasise the need to identify and prioritise the costs and benefits that have the greatest impact in complex infrastructure projects. Nevertheless, according to CBA, the project with the highest estimated monetary (economic) value will be selected. In contrast, the premise behind input–output tables is that infrastructure decisions are taken to maximise GDP (i.e. gross domestic product) growth (Ploszaj *et al*., 2015).

Both techniques are useful for analysing the economic value of infrastructure, but their different approaches dictate which of the tools should be used. CBA relies on expert evaluation of an infrastructure project's specific impacts by a multidisciplinary team of experts to estimate the construction, economic, environmental, and social value over its lifetime (European Commission, 2014). Conversely, input–output tables model the flow of economic value through an economy (Giesekam *et al*., 2018). An input–output table therefore shows the effect of new infrastructure on every sector of the economy and thus reveals dependencies between sectors. The inputs and outputs are generated by predetermined equations, meaning that input–output analysis is potentially less biased than the experts' approach used by CBA. New infrastructure impacts the variables that feed into the predetermined input–output equations. In other words, the benefit of new infrastructure is modelled by how it impacts the economic value across all sectors, hence naturally identifying and (where possible) capturing all forms of value generation and, critically, revealing the value-generating opportunities offered by infrastructure interdependencies – a necessary feature of 'whole-system' approaches to value generation.

CBA and input–output tables thus provide analysis from opposite perspectives. CBA employs a bottom-up perspective in which it aggregates individual impacts of the infrastructure into a singular economic value. Input–output analysis employs a top-down perspective in which it attributes anticipated changes throughout the entire economic system to the ripples emanating from changes in infrastructure due to a 'system intervention' (i.e. the introduction of new infrastructure). To combine and study the singular economic values of different infrastructure projects estimated using CBA requires that the starting points (the baseline for the analysis) are similar, for example same assumptions, same costs and benefits, and so on. On the other hand, the holistic equations used by the input–output tables are fixed, meaning that the starting point (for the top-down analysis) is the same. To conclude, input–output tables are more useful than CBA when studying infrastructure interventions as a whole, as this research does, because they compare similar and less subjective data, whereas CBA works better for smaller infrastructure projects when specific impacts can be assessed.

This research analyses the economic impact of infrastructure using economic input–output tables, because, whether correctly or not, GDP is viewed as a measure of economic progress, which is 'the value of the goods and services produced by the nation's economy less the value of the goods and services used up in production' (Dynan and Sheiner, 2018).

### 3.2 Critical summary of cost–benefit analysis

CBA was developed in the mid-nineteenth century and has substantially evolved to its current form (Saad and Hegazy, 2015). CBA assigns a monetary value unit to every impact of infrastructure (Vickerman, 2007). By the turn of the twenty-first century, CBA was stated to be the most prevalent economic value assessment methodology for infrastructure projects (Hayashi and Morisugi, 2000) largely because it was, again rightly or wrongly, considered to provide a summary of all the benefits and costs of a project (Mouter, 2014). CBA has project-specific applications and varies in the types of costs and benefits measured, quantification methods, time periods considered, and discount factors used, according to Couture *et al*. (2016). The key features are as follows:

- Benefit: Benefit is a utility gain and is measured by the individual's willingness to pay or willingness to accept compensation, meaning the demand (Couture *et al*., 2016). Benefits are typically assessed using the additional productivity provided by the project (Vickerman, 2007).
- Cost: Cost can be considered as a 'well-being loss', which is measured by the individual's willingness to accept/tolerate loss or willingness to pay to prevent loss (Couture *et al*., 2016).
- Time period (Couture *et al*., 2016; Van Wee, 2007): Costs and benefits occur at different time intervals within the CBA timeline. Future costs and benefits are discounted to define the present value of money (or net present value, NPV). The present value of benefits and costs are then used to determine the NPV of projects, which also allows for comparison. The discount rate cannot be easily estimated and is usually based on historical data.
- Sensitivity: Sensitivity analysis is used to estimate the uncertainty of project parameters (discount rate, volume of consumption, and so on) that are used in the CBA analysis (Couture *et al*., 2016). The sensitivity models are error-free if they do not reveal large unexpected changes in NPV (Couture *et al*., 2016). Sensitivity analysis is challenging as the number of projects increases, since the number of assumptions increases too. Moghayedi and Windapo (2023) showed that while it is challenging to assess cost uncertainties even in linear infrastructure projects, sophisticated models (e.g. AI) can lead to more accurate cost estimations and better management of financial risks.

The above analysis means that using CBA is challenging when studying the infrastructure system as a whole, as this research does.

#### 3.2.1 Advantages and disadvantages of cost–benefit analysis

CBA can include a wide range of infrastructure costs and benefits, provided that there is sufficient information about them (Hoogmartens *et al*., 2014). The primary weakness of CBA is its dependency on complete information. When values do not have a monetary (market) price, CBA attempts to assign a monetary value (Ackerman and Heinzerling, 2002). The standard CBA methods for estimating monetary value are based on survey data and secondary data (e.g. 'equivalent' consumer preferences; Atkins *et al*., 2017). The challenge, again, is that many environmental and social values are either problematic or impossible to quantify (and indeed are claimed by some to be entirely subjective; see Ackerman and Heinzerling, 2002), and the uncertainty of CBA increases.

The uncertainty of CBA also increases its vulnerability to bias (e.g. political bias) to create support for an infrastructure project (Van Wee, 2007). Flyvbjerg (2007) states that many large infrastructure projects that are supported by CBA are later discovered to have higher than expected costs and lower than expected benefits. A comprehensive CBA analysis, by a multidisciplinary team, is required to determine whether there is sufficient information to reliably quantify the impacts of the project (European Commission, 2014; Zhuang *et al*., 2007).

In contrast, CBA is less susceptible to the subjectivity of factor weights, since it compares different alternatives for the same infrastructure project under the same framework (Van Wee, 2007). Likewise, Branigan and Ramezani (2018) state that CBA ensures that the same criteria are applied when evaluating the costs and benefits of different alternatives of the same infrastructure and reduces bias due to information asymmetry.

Furthermore, CBA entails a considerable number of assumptions, which are considered acceptable and negligible at small scales; however, for large infrastructure projects, these assumptions amplify and can affect the assessment (Vickerman, 2007). This demonstrates that CBA is not suitable for studying large infrastructure systems. As Vickerman illustrates, the assumptions include predicted and static demand (even when accounting for income rises and technology advancements), fixed dependencies between infrastructures (meaning that new infrastructure is equally used with the existing one), and perfect competition among infrastructure providers (Vickerman, 2007). These assumptions increase the uncertainty of the CBA process and contradict the social value of infrastructure, which acknowledges that the demand varies based on specific factors (e.g. new infrastructure provides better quality of services).

CBA is susceptible to changes in discount rates and to the time period considered. The discount rates are estimated based on historical data and the analyst's interpretation, implying that the discount rates are influenced by potential biases. Analysts tend to favour infrastructure with short-term benefits (aligning with shareholder capitalism) without considering the latest definitions of infrastructure, where infrastructure is a long-term resource (aligning with stakeholder capitalism).

Saad and Hegazy (2015) found that CBA does not support multi-level decisions, such as the allocation of value between different infrastructure systems, an argument that is consistent with the contention that CBA does not consider wide economic impacts. This might be illustrated by a proposal for a new section of railway connecting a number of local centres (e.g. the reinstatement of the Ivanhoe Line in Leicestershire; Scott Wilson, 2009). A traditional CBA might primarily focus on direct costs and benefits, such as construction expenses, ticket revenues, and travel time savings for passengers, all based on reasonable assumptions and historical experience (e.g. on modal shift). However, this approach can overlook several broader economic impacts. For instance, while the rail system might stimulate economic growth in less developed areas by enhancing access to markets and employment opportunities, any assumptions made here are likely to be far less certain; the benefits to the physical health and mental health (or well-being) of those members of the public that are affected is more problematic still to quantify and monetise, yet this social benefit could prove very considerable indeed. Likewise, while the CBA might account for immediate environmental costs, such as land use and emissions during construction, it might not fully consider the manifold long-term environmental benefits of reduced car travel, a focus of new housing development away from car-dependent rural areas, improved freight logistics, reduced impacts on biodiversity, and so on. Moreover, the allocation of value between different places, communities, and timescales can be complex. Table 2 summarises the various advantages and disadvantages of the CBA as documented in the literature.

### 3.3 Critical summary of input–output tables

Input–output tables use a pre-existing system of equations to avoid some of the biases, such as political bias, to which CBA is potentially prone. This is because the input–output tables are developed in a consistent manner across the world. The input–output tables show the value flow through an economy by relating the demand in economic sectors to the required inputs from other sectors, thus enabling the study of dependencies between sectors.

Input–output tables were initially developed in the seventeenth century and later formalised by Wassily Leontief in 1974. The input–output tables are $n \times n$ matrices that include all the economic sectors of an economy and the value flow between these sectors. The rows of the input–output tables show the output of each economic sector, and the columns show the input to each sector. Each cell of the input–output table is the output of a specific sector to another sector (input). Input–output tables aim to 'model linkages between the productive sectors of an economy' (Avelino and Dall'erba, 2018), which inherently focus on large-scale and large-impact economic events (e.g. infrastructure).

Table 2. Advantages and disadvantages of cost–benefit analysis (Kalyviotis, 2022)

| Advantages | Disadvantages |
| --- | --- |
| Previous historical data to infer from costs and benefits are already theoretically known in infrastructure projects | Cost–benefit analysis tends to omit environmental and social benefits of projects |
| Cost–benefit analysis is neutral when compared with other multi-criteria analysis | Requires lots of assumptions: risk of errors for large-scale additions |
| Provides direct comparison between alternatives. Cost–benefit analysis allows easy comparison of different options for same project, reducing bias and improving clarity | Often subjected to optimism bias |
| Incorporates specific elements of projects, including non-economic externalities | Does not support multi-level decisions |
| | Many project impacts are difficult to quantify |
| | Cost, time, and expertise to conduct a rigorous CBA can be extensive. As such, this should be scaled to the size of the civil infrastructure. Less effective as the project gets bigger |
| | The value of time or some other non-quantifiable aspects used to evaluate the cost and benefit could lead to a biased result if manipulated wrongly |
| | Data constraints and uncertainty with large projects with long-term horizons: CBA may not be able to account for inflation, interest rates, changes to cash flow, and the present value of money. In addition, uncertainty in the forecasts for future revenue or sales, expected costs and cash flows, or event that impacts of climate change may limit the actual performance of the CBA |
| | Also, wrongly determining the discount rate will lead to a skewed results |
| | Variations in scope, lack of consistency of CBA based on regional guidelines. This can lead to different outcomes that ultimately impact the evaluation of a decision |
| | Equity is not considered CBA since benefits or costs for one stakeholder is not given greater value than others |

Input–output tables are comprehensive tables that provide the inter-industry transaction statistics of a specific region or an economy (Avelino and Dall'erba, 2018; Jun et al., 2018). They provide the total demand and supply of each economic sector and a way to assess the correlation magnitude between various economic sectors (Jun et al., 2018), which is the main objective of this research.

3.3.1 Advantages and disadvantages of input–output tables
The information used in input–output tables is usually provided by government agencies in collaboration with international financial institutes, and thus the transparency is ensured and guaranteed (Wang and Charles, 2010). This implies that the information used in input–output tables as a methodology is ideologically neutral, as it does not allow for 'specific behavioural conditions for the individual, companies, or indeed, the state' (Wang and Charles, 2010). The input–output tables include quantitative data, which can be difficult to obtain (Worth et al., 2007).

The main advantage of input–output tables is that they explain the impact of entire sectors on others and thus represent the 'interconnectedness of economic processes' (Wang and Charles, 2010). Input–output tables enable the assessment of the economic impact caused by changes in any sector on the overall economy of an area (Jun et al., 2018). Changes to the infrastructure system (e.g. recessions, expansions, or emergencies) are considered a shock in input–output analysis (Zhao and Kockelman, 2004), 'shocks' being changes that impact the input–output system of equations that model a sector's production and demand for resources (Yu, 2017). Although input–output analysis focuses on shocks/changes, input–output tables can also be used to describe the current situation of a system, as this research does. Each sector or change in a sector (e.g. infrastructure systems) has direct and indirect economic effects on the other sectors. The direct effect is the direct output of this sector, while the indirect effect is how this sector induces production factors to increase economic value in other sectors. Input–output tables have low data requirements for estimating the economic value of a sector (Avelino and Dall'erba, 2018) and are graphic and easy to use, while being readily accessible (Wang and Charles, 2010). The accessibility of the data makes input–output tables a cost-effective methodology.

This research aims to model infrastructure dependencies during typical situations, and not for emerging situations. A well-calibrated input–output model can reveal systematic economic benefits that might not be evident to experts conducting a CBA.

This research develops a new model to capture the dependencies of critical infrastructure. Once the underlying assumptions of the tables are understood, it is straightforward to explain the results of input–output tables to decision makers (Ploszaj *et al*., 2015).

The simplicity of the input–output tables, which enables decision makers to understand them, is also their main disadvantage. Input–output tables are clearly transparent in the way they model the economic value of infrastructure on interconnected economic sectors (Ploszaj *et al*., 2015) because they separate out environmental and social value, that is a piece of infrastructure may have a similar effect on GDP, but vastly different social and environmental values.

The second weakness of the input–output tables is that they are not suitable for individual infrastructure projects (Gkritza *et al*., 2007). Developing an input–output table and their economic dynamics requires extensive economic data and validation (Yu, 2017). Many organisations or even national governments lack the capacity to develop input–output tables, especially when infrastructure projects have a wide impact on the economy (Duncombe and Wong, 1998). This research focuses on infrastructure as a whole and not on individual infrastructure projects. Moreover, the input–output tables were already provided by the World Input-Output Database for the UK. Therefore, this disadvantage does not affect the analysis in this research.

The most critical assumption of traditional input–output tables is that the economy is demand-driven (Van Wee, 2007) without considering technological changes, price changes, and international trade pattern changes (e.g. COVID-19 pandemic). Prices are constant and not subject to change (Van Wee, 2007), making the input–output tables a fixed price model. The static nature of input–output tables is a result of the over-generalisation of this methodology (Onat *et al*., 2015). This is illustrated by the lack of account of 'long-term economic, industrial, and demographic changes' (Gkritza *et al*., 2007). Thus, input–output tables should be updated periodically over time. The UK has published only a limited number of input–output tables, which compelled the authors to look for other sources.

The final significant limitation is that industries belonging to the same economic sector produce one homogeneous product. For instance, if a construction company (construction sector) owns and maintains (manufacturing sector) its own truck fleet to provide a transportation service, such as transporting precast structures (transport sector), then the economic value is attributed to the construction sector. This research focuses on infrastructure, so it clarifies the boundaries of which parts of economic sectors are regarded as infrastructure. Table 3 summarises the various advantages and disadvantages of the input–output tables as documented in the literature.

### 3.4 Comparison of methods and conclusions

CBA and input–output tables use different tools and perspectives to analyse the economic impact of infrastructure. The authors considered comprehensiveness as the first criterion. CBA can be applied when the infrastructure investment creates quantifiable (and monetisable) social and/or environmental value, whereas input–output tables, because they analyse the interrelation between different sectors of the economy, provide a more comprehensive and insightful view of an infrastructure's economic impacts. Therefore, to evaluate the full range of the value of infrastructure systems that encompass the considerations of economic, social, and environmental well-being, input–output tables should be combined with other databases.

A second required criterion is the ability of the methods to compare scenarios or alternatives. While both approaches have distinct advantages for analysing different scenarios, input–output tables are particularly helpful in that they enable interpretations when infrastructure induces economic growth in different sectors of the economy. In contrast, CBA's valuing of infrastructure in monetary terms, which allows direct comparisons to be made from the subset of topics considered, nevertheless results in uncertainty regarding how the economy is influenced by infrastructure.

Table 3. Advantages and disadvantages of the input–output tables (Kalyviotis, 2022)

| Advantages | Disadvantages |
| --- | --- |
| Allows user to understand impact of certain industries on others, and shows interconnectedness of economic processes | Large uncertainties due to generalisation<br>• Uncertainty of results<br>• Lack of consensus on a preferred model leads to a decrease in its credibility |
| When compared with other methodologies; lower data requirements, increased ease of use, good transparency. Easy to explain results to decision makers | This is a static model |
| Allows user to take into account factors that are not easily quantifiable | No specialised methodologies to validate the data used in input–output tables |
| More cost effective | Input–output models do not consider any productive constrains of the economy. Does not attempt to model externalities |
| Measure the economic impact of infrastructure on the overall economy | It is difficult to develop accurate underlying assumptions and equations, so it should be kept simple |

The specific differences discussed above indicate the advantages that input–output tables (top-down approach) have over CBA (bottom-up approach) when studying infrastructure dependencies. However, it should be emphasised that input–output tables and CBA are not mutually exclusive; when there is insufficient knowledge about the impacts of infrastructure, each method can supplement the other. Both techniques can be applied to analyse infrastructure decisions, though with differential efficacy under certain conditions: CBA is optimal when there is sufficient information to calculate a monetary value for all economic, social, and environmental values of an individual infrastructure project, while input–output tables are most effective when the environmental and social values of infrastructure are distinguished from the economic value. That way, the value generated in the input–output tables due to infrastructure is a true representation of the infrastructure's aggregate economic value (Wang and Charles, 2010). Moreover, input–output tables are superior to CBA when the economic impacts of infrastructure are spread across many economic sectors or other infrastructures.

## 4. Results and analysis

The importance of interdependencies between infrastructure systems has often been overlooked in past research (Kalyviotis, 2022). This current study uses document analysis as the source of empirical data and conducts secondary data analysis, which by definition is the analysis of pre-existing data (Heaton, 2000: p. 1). Administrative records and, more specifically, symmetric (product-by-product) input–output tables reveal past dependencies by providing estimates of domestic and imported products (product-by-product tables are also known as industry-by-industry tables in the US) to intermediate consumption and final demand and associated multipliers (Peters *et al*., 2008). These tables were used to obtain part of the empirical data and achieve the objective of this research.

The World Input-Output Database provided the economic value dependencies between different sectors. These documents produced numerical findings. In line with the numerical findings of the World Input-Output Database, this study uses quantitative data collection to provide a detailed description and explanation of how value is created in interconnected infrastructure systems. The documents targeted investors and the public audience, so they lacked scientific or economic data/text to support them. These documents were knowledge drivers and established the guidelines for the processes of the public organisations, facilitating their function and development through the same processes as a whole. These documents were generated from different day-to-day or month-to-month reporting systems over the period 2000–2014. Therefore, the research strategy followed is archival. Archival research refers to the analysis of administrative records and documents as the main source of data because they are products of day-to-day activities (Saunders *et al*., 2009: p. 587). The documents comprise a variety of written and visual material that represents the values of the related organisations and their existing dependencies. The researcher was aware that these manuals were created for a different purpose than the aim of the current research (Olson, 2010: pp. 319–320). Research in archival documents cannot predict the outcome and its nature is iterative (Hill, 1993: p. 6). Archival data usually help the researcher to identify the true nature of complex networks and reveal unrecognised human interactions. The relationship between the literature presented and the archival documents is reciprocal (Hill, 1993: p. 62). According to Hill (1993), there is no fixed archival analysis method and the author learns in the process how to extract information. The authors decided to implement process analysis networks and cohorts due to the research nature. This type of analysis uses tables, diagrams, models, and networks of interactions along with organisational linkages (Hill, 1993: p. 62) and aligns with the network theory that is used for infrastructure systems modelling (Dunn *et al*., 2013). The steps for analysing the documents are as follows: (1) reading the documents, recognising and highlighting linkages with the research proposition; (2) creation of networks and/or tables with data needed; and (3) mapping economic value interdependencies.

Following the three-step process analysis networks and cohorts (Hill, 1993), step 1 involved the use of symmetric (product-by-product) input–output tables, which comprise product input–output groups (IOGs). The research proposition requires an industry-based analysis focusing on transport, energy, waste, communication, and water. Each of the IOGs was categorised according to their main product or service as transport, energy, waste, communication, water, or other goods/services. Step 2 involved the creation of tables with the empirical data discussed above. Step 3 involved the mapping of the economic value interdependencies (see Table 4, the values for which were determined using Equation 1), and the map was used for the development of a mathematical model (function).

The World Input-Output Database (2018) provided part of the empirical data to meet the objectives of this research. Created by Timmer *et al*. (2015, 2016) the International Standard Industrial Classification revision classified the 56 sectors (Timmer *et al*., 2016). The input–output tables used were balanced with the RAS method (Timmer *et al*., 2015), the most widely known and commonly used automatic procedure for this procedure (Trinh and Phong, 2013: p. 135).

### 4.1 Principal component analysis

Input–output tables can be analysed using regression analysis (see Zhang *et al*., 2021a, 2021b), considering the transport sector the as the dependent variable and energy, waste, communication, and water sectors as independent variables, to provide preliminary insights into which of the sectors are significantly related. This regression analysis demonstrated that there was multicollinearity, the presence or absence of which is shown by variance inflation factor (VIF; Field, 2009: p. 224) – multicollinearity does not occur when VIF < 10 (Myers, 1990). Multicollinearity is the condition where one or more of the predictor variables in a multiple regression model can be almost precisely estimated from the others using linearity (Field, 2009: p. 224). In this situation, the model is unstable since the coefficients may vary significantly in response to small changes in the model or the data (Field, 2009: p. 273 and 297).

**Table 4.** Economic infrastructure interdependencies

| Transport | Energy | | Water | Communication | Waste | | |
|---|---|---|---|---|---|---|---|
| | Electricity, transmission, and distribution | Gas; distribution of gaseous fuels through mains; steam and air conditioning supply | Natural water; water treatment and supply services | Telecommunications services | Sewerage services; sewage sludge | Waste collection, treatment and disposal services; materials recovery services | Remediation services and other waste management services |
| Land transport | | | | | | | |
|   Motor vehicles, trailers and semi-trailers | ✓ | ✓ | ✓ | ✓ | ✓ | ✓ | ✕ |
|   Rail transport services | ✓ | ✓ | ✓ | ✓ | ✓ | ✓ | ✕ |
|   Wholesale and retail trade and repair services of motor vehicles and motorcycles | ✓ | ✓ | ✓ | ✓ | ✓ | ✓ | ✕ |
|   Land transport services and transport services by way of pipelines, excluding rail transport | ✓ | ✓ | ✓ | ✓ | ✓ | ✓ | ✓ |
| Water Transport | | | | | | | |
|   Ships and boats | ✓ | ✓ | ✓ | ✓ | ✓ | ✓ | ✕ |
|   Water transport services | ✓ | ✓ | ✓ | ✓ | ✓ | ✓ | ✕ |
|   Repair and maintenance of ships and boats | ✕ | ✕ | ✕ | ✕ | ✓ | ✓ | ✕ |
| Air Transport | | | | | | | |
|   Air transport services | ✓ | ✓ | ✓ | ✓ | ✓ | ✓ | ✕ |
|   Air and spacecraft and related machinery | ✓ | ✓ | ✓ | ✓ | ✓ | ✓ | ✕ |
|   Repair and maintenance of aircraft and spacecraft | ✕ | ✕ | ✕ | ✓ | ✕ | ✕ | ✕ |
| Other transport equipment | ✓ | ✓ | ✓ | ✓ | ✓ | ✓ | ✕ |

Principal component analysis (PCA) was applied to reduce this correlation. Energy, waste, communication, and water were examined without dividing them into their IOGs; the other sectors comprised 43 correlated IOGs. The sum of the variables was then transformed into uncorrelated factors using PCA (Field, 2009: pp. 633–638). PCA explores the linear combination of correlated variables to account for the maximum variability of variables by converting them to linear uncorrelated factors (Field, 2009: p. 638) using eigenvalues. This enables the transformation of the previous equation into an equation without multi-collinearity. Eigenvalues, also known as characteristic roots, characteristic values (Hoffman and Kunze, 1971: p. 182), proper values, or latent roots (Marcus and Minc, 1988: pp. 144–145), provide the factors by which the axes (i.e. the direction of the axes) change for a linear transformation (compression). First, the eigenvalues are determined, allowing the authors to calculate the number of factors that can substitute the variables (see the Video online).

As shown in Table 5, three factors account for 90.7% of the variables (89.8% of the rotated) and four new factors account for 93.5% of the variables (even the rotated). Then four new factors will be applied. The main issue with these factors is that they cannot be based on the sectors of interest, as they will have a new unit and it will not be feasible to calculate the values of the sectors of interest easily. Instead of applying the new factors, the components of each factor were examined and the highest component selected considering the sectors of interest. Thus, the components had the same unit as the sectors of interest before the PCA.

The objective is to develop a model that incorporates energy, waste, water, and communication and/or at least a model that combines the same measurement with them, namely economic demand. PCA results have virtual measurement units, not economic demand. To develop a model that includes energy, waste, water, and communication, the four sectors of interest were used to see which of the four factors correspond with them and then IOGs that cover most of the missing correlation were selected (see Appendix A online).

It is obvious that the correlation cannot be reduced (VIF > 10) regardless of the combination of IOGs, since energy, waste, water, and communication all majorly affect the first factor. The final model should include the sectors of interest: energy, waste, water, and communication. It is clear that all the sectors of interest mainly affect the first factor. The second factor is mainly affected by printing and reproduction of recorded media (first in the second factor both for component matrix and rotated component matrix). The third factor is mainly affected by manufacture of textiles, wearing apparel, and leather products (first in the third factor both for component matrix and rotated component matrix). The fourth factor is mainly affected by mining and quarrying (first in the fourth factor for component matrix and third for rotated component matrix).

Using the above factors for developing the linear model, the correlation has significantly reduced. The new model explains 98.3% of the variance of the data:

1. $$\begin{aligned}Y_{cr} = &\ 1.299\,X_{cr1} + 1.739\,X_{cr2} + 1.687\,X_{cr3} \\ &+ 1.735\,X_{cr4} + 1.947\,X_{cr5} + 5.131\,X_{cr6} \\ &+ 0.546\,X_{cr7} + 30507.785\end{aligned}$$

where $X_{cr1}$: value added from energy, $X_{cr2}$: value added from waste, $X_{cr3}$: value added from communication, $X_{cr4}$: value added

Table 5. Total variance explained

| | Initial eigenvalues | | | Extraction sums of squared loadings | | | Rotation sums of squared loadings | | |
|---|---|---|---|---|---|---|---|---|---|
| **Component** | Total | % of variance | Cumulative % | Total | % of variance | Cumulative % | Total | % of variance | Cumulative % |
| 1 | 30.931 | 67.241 | 67.241 | 30.931 | 67.241 | 67.241 | 27.307 | 59.363 | 59.363 |
| 2 | 6.599 | 14.346 | 81.587 | 6.599 | 14.346 | 81.587 | 7.314 | 15.900 | 75.264 |
| 3 | 4.190 | 9.108 | 90.695 | 4.190 | 9.108 | 90.695 | 6.666 | 14.490 | 89.754 |
| 4 | 1.307 | 2.842 | 93.537 | 1.307 | 2.842 | 93.537 | 1.740 | 3.783 | 93.537 |
| 5 | 0.876 | 1.904 | 95.441 | — | — | — | — | — | — |
| 6 | 0.612 | 1.331 | 96.772 | — | — | — | — | — | — |
| 7 | 0.441 | 0.959 | 97.730 | — | — | — | — | — | — |
| 8 | 0.374 | 0.813 | 98.544 | — | — | — | — | — | — |
| 9 | 0.215 | 0.468 | 99.012 | — | — | — | — | — | — |
| 10 | 0.160 | 0.347 | 99.359 | — | — | — | — | — | — |
| 11 | 0.120 | 0.260 | 99.619 | — | — | — | — | — | — |
| 12 | 0.090 | 0.197 | 99.816 | — | — | — | — | — | — |
| 13 | 0.056 | 0.121 | 99.937 | — | — | — | — | — | — |
| 14 | 0.029 | 0.063 | 100.000 | — | — | — | — | — | — |

Extraction method: principal component analysis from water, $X_{cr5}$: value added from printing and reproduction of recorded media, $X_{cr6}$: value added from manufacture of textiles, wearing apparel, and leather products, and $X_{cr7}$: value added from mining and quarrying. This last exemplar model is not representative of the total value as it does not take into account a lot of sectors, but it is a good display of the interdependencies.

## 5. Conclusions

This research discusses the advantages of input–output tables (a top-down approach) over CBA (a bottom-up approach) for studying infrastructure dependencies. CBA works best when there is adequate information to calculate a monetary value for all economic, social, and environmental impacts of an individual infrastructure project. Input–output tables are most effective for analysing the economic value when the environmental and social impacts of infrastructure are clearly distinguished from the economic impact. Input–output tables work better than CBA when the economic impacts of infrastructure are dispersed across many economic sectors or other infrastructures. Moreover, CBA is inadequate to assess the economic impact of infrastructure on the overall economy.

This study underscores the critical importance of recognising past dependencies in infrastructure research, which has often been overlooked. By employing document analysis and secondary data analysis, the research effectively utilised symmetric input–output tables and the World Input-Output Database to reveal economic value dependencies across various sectors. The archival research strategy, focusing on administrative records and documents, provided a comprehensive understanding of the interconnectedness within infrastructure systems.

The use of PCA addressed multi-collinearity issues, enabling the development of a robust model that explains a significant portion of the variance in the data. The final model, incorporating key sectors such as energy, waste, water, and communication, highlights the intricate economic interdependencies and offers valuable insights for future infrastructure planning and policy-making.

The main finding, which addressed the main question of this research, is the determination of the value dependencies between the transport infrastructure and the other four economic infrastructures. The economic value was investigated considering the demand for the transport sector in the UK and how this demand interacts with the demand for energy, water, communications, and waste. This study was conducted from an engineering perspective, meaning that some industries considered in the manufacturing sector were considered as part of the transport sector, and approaches such as the input–output tables were used. The World Input-Output Database provided economic data for linear regression, statistical analysis, and PCA, well beyond the input–output tables determined from the Office for National Statistics. The analysis of the economic data provided the required data for a generalised model, and an economic linear model was developed. The economic transport infrastructure interdependencies, based on the demand between sectors, were identified showing that transport infrastructure is economically complemented by the energy, communication, water, and waste infrastructures.

Overall, this research contributes to a deeper understanding of how value is created and sustained in interconnected infrastructure systems, emphasising the necessity of considering historical dependencies to inform contemporary and future infrastructure development.

## Acknowledgements

The authors gratefully acknowledge the University of Birmingham, the University of Illinois at Urbana–Champaign, and the University of Crete and the financial support of the UK Engineering and Physical Sciences Research Council under grant numbers EP/K012398 (iBuild: Infrastructure Business Models, Valuation and Innovation for Local Delivery), EP/J017698 (Transforming the Engineering of Cities to Deliver Societal and Planetary Wellbeing, known as Liveable Cities), and EP/R017727 (UK Collaboratorium for Research on Infrastructure and Cities).